# Direct Experimental Evidence of the Statistical Nature of the Electron Gas in Superconducting Films

**Mario Iannuzzi, Massimiliano Lucci, Ivano Ottaviani**

Physics Department, University of Rome "Tor Vergata", Rome, Italy
Email: iannuzzi@roma2.infn.it

## Abstract

In an Nb film an alternate electrical current is partitioned at a *Y*-shaped obstacle into two splitted beams. The intensity-fluctuation correlation of the two beams (cross-correlation) and the intensity-fluctuation correlation of one beam (auto-correlation) are measured within a low-frequency bandwidth as a function of the incident beam intensity, at temperatures $T$ above or below the temperature $T_c$ of the superconductive transition. The results of these measurements reveal the statistical nature of the electron gas in the normal film and in the superconducting film. The conceptual scheme of the present experiment is a version of the Hanbury Brown and Twiss (HBT) experiment, here adopted for a gas of particles in a solid.

## Keywords

## 1. Introduction

The Cooper pairs, which play the fundamental role in the BCS theory of low-temperature superconductivity, are generally described, although with explicit caution, as bound pairs of electrons in a highly coherent state; the formation of few pairs encourages the formation of others in a cooperative way, a tendency quite analogous to what happens in a Bose condensation. For instance, with Feynman's wording [1]: "The two electrons which form the pair are not like a point particle… They are really spread over a considerable distance… The mean distance between pairs is relatively smaller than the size of a single pair. Several pairs are occupying the same space at the same time… We will accept, however, the idea that electrons do, in some manner or other, work in pairs… behaving more or less like particles. Since electron pairs are bosons, when there are a lot of them in a given state there is an especially large amplitude for other pairs to go to the same state. So nearly all of the pairs will be locked down at the lowest energy in exactly the same state…".

Now, such a general and qualitative description of the physical nature of the Cooper pairs being universally taken as true, it would be nice to verify it by a direct experimental test. Such a test is feasible, as we discuss in the present letter describing the experiment we have performed.

First we have argued that the physical nature (bosonic, fermionic, or classical) of a beam of particles can effectively be distinguished by measuring its temporal intensity-correlation function. In this regard we may recall that Hanbury Brown and Twiss (HBT) [2] long ago measured the bunching effect of a beam of optical photons, so showing their bosonic nature; and more recently the antibunching effect of fermions was observed in a beam of free non-interacting neutrons [3]. Other fermionic and bosonic systems have also been studied [4] [5]. These results have convinced us that a HBT experiment might bring directly to light the statistical properties of a beam of Cooper pairs, and have motivated the measurements presented below in this Letter.

## 2. Method

Let $I(t)$ be the time-dependent current intensity of a beam of charged particles, and $\Delta I$ it's fluctuation around the average intensity $\langle I \rangle$, due to the fluctuation $\Delta N$ of the number of particles $N$ detected within a certain time interval. A classical stream of carriers originating from an equilibrium reservoir will obey Poisson statistics with $\langle (\Delta N)^2 \rangle = \langle N \rangle$, whereas a stream of Bose carriers will obey Bose-Einstein statistics with $\langle (\Delta N)^2 \rangle$ larger than $\langle N \rangle$, and a stream of Fermi carriers will obey Fermi-Dirac statistics with $\langle (\Delta N)^2 \rangle$ smaller than $\langle N \rangle$. Now, along the lines of the HBT pioneering experiment on optical photons, let an experiment on quantum carriers, instead of measuring the intensity fluctuations of a single beam, measure the intensity fluctuation correlations $\langle \Delta I_a \Delta I_b \rangle$ between two partial beams, $I_a$ and $I_b$, originating from a beam splitter. General expressions giving the theoretical cross correlation $\langle \Delta I_a \Delta I_b \rangle$ can be obtained from the cross correlation function $\langle \Delta n_a \Delta n_b \rangle$ of the occupation numbers, $n_a$ and $n_b$, of the two partitioned beams, as it is shown in [6] for beams emitted from an equilibrium reservoir ( of bosons or fermions ).

In the presence of a steady current, provided that $eV \gg K_B T$, the cross correlation for transport fluctuations of the two beams $I_a$ and $I_b$, in small conductors and in the limit of low frequencies, has been derived theoretically as [6]:

$$\langle \Delta I_a \Delta I_b \rangle = \pm 2qI\Delta\nu t_a t_b \qquad (1)$$

where $t_i (i=a,b)$ is the transmission probability of the portioned beam-$i$, $I$ is the total current intensity (for an alternate current we adopt the notation $I = I_{\text{eff}}$), $q$ is the carrier charge ($q$ = e for fermions and $q$ = 2e for Cooper pairs), and $\Delta\nu$ is a flat frequency band-width within the low-frequency spectrum of the current fluctuations. Here a small electrical conductor is a conductor of small resistance, such that the wavelike transport of carriers preserves phase coherence over distances larger than the distance between the two contacts of the beam partition. With $a = b$ (auto-correlation), the sign of the correlation is always positive, both for fermions and bosons, and formula (1) reduces to the standard expression for shot noise $\langle (\Delta I)^2 \rangle = 2eI\Delta\nu$. With $a \neq b$ (cross-correlation), for fermions the correlation is always negative. For bosons, even though as a rule the correlation would be negative because flux conservation requires that an increase at one partition must be compensated by a decrease at the other partition, yet there are many circumstances under which the sign of the cross-correlations is positive, among them, the conductors with one of the two terminals partitioned into two leads, like our Nb samples [6]. Experimental evidence of the anticorrelation predicted by formula (1) for a gas of fermions was obtained on a beam of electrons in the quantum Hall regime [7].

## 3. Experiment and Results

Let us consider now superconducting films. Obviously, expression (1) with negative sign applies also to such films at temperatures above the critical temperature $T_c$ (Fermi gas); differently, at temperatures below $T_c$ the Cooper pairs should behave like bosons with consequent positive cross-correlation at a *Y*-shaped partition. Yet, the fluctuation correlations cannot be discussed without addressing the cooperative effect of such pairs analogous to the Bose condensation transition: therefore, now the incident state (*i.e.* the state arriving at the partition obstacle ) is a single state of energy and momentum containing a precise number of many particles (nearly all of the pairs), and the fluctuations $\Delta I_a$ and $\Delta I_b$, which are a consequence of the probabilistic transmission either in partition *a* or *b*, will have opposite signs if measured at the same time. With $n_I$ as the occupation number of the incident state containing a precise number of particles, and $n_a$, $n_b$ as the occupation numbers of the transmitted states, the cross-correlation is $\langle \Delta n_a \, \Delta n_b \rangle = -n_I t_a t_b$, and the auto-correlation is $\langle (\Delta n_a)^2 \rangle = \langle (\Delta n_b)^2 \rangle = n_I t_a t_b$.

This yields:

$$\langle \Delta I_a \, \Delta I_b \rangle = -qI\Delta v t_a t_b = -2eI\Delta v t_a t_b \tag{2a}$$

$$\langle (\Delta I_a)^2 \rangle = \langle (\Delta I_b)^2 \rangle = 2eI\Delta v t_a t_b \tag{2b}$$

Consequently, the expectation value of the correlation is still given by expression (1), with $q = e$.

An experiment aimed at enlightening the statistical nature of the Cooper pairs will therefore measure the cross correlation $\langle \Delta I_a \, \Delta I_b \rangle$ above and below $T_c$, testing the validity of such predictions.

We have performed an experiment whose general scheme is represented in **Figure 1**. The correlator outcome is a voltage $V_C$ proportional to $\langle \Delta I_a \, \Delta I_b \rangle$. The measurements have been carried out, at different temperatures above and below $T_c$, on various specimens of Nb films of small resistance, $R \leq 25\Omega$ at $T = 293$ K. The particle source was a current source, and the two exit terminals of the *Y*-partition, virtually grounded, were connected to a current-voltage converter which allowed to measure separately $I_a$ and $I_b$. The width of the pass-band filters was $\Delta\nu$ = 300 Hz - 600 KHz for the all measurements.

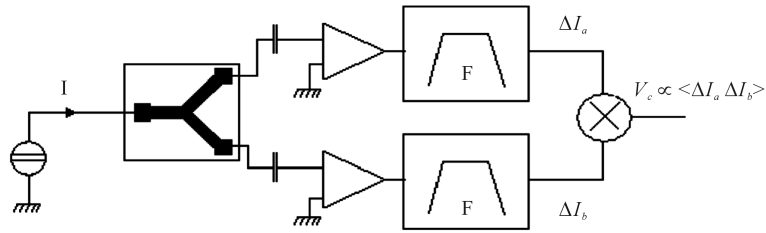

**Figure 1.** Scheme of the experiment on Nb films. The frequency of the incident beam I was 3 MHz. The total length of all Y-shaped specimens was 7 mm, their cross section $\leq 400 \times 0.3$ um$^2$, and their electrical resistance at room temperature $\leq 25$ Ohm. The pass band of the filters F was in the range 300 Hz to 600 KHz. Correlator readings $V_C$ are proportional to $\langle \Delta I_a \, \Delta I_b \rangle$.

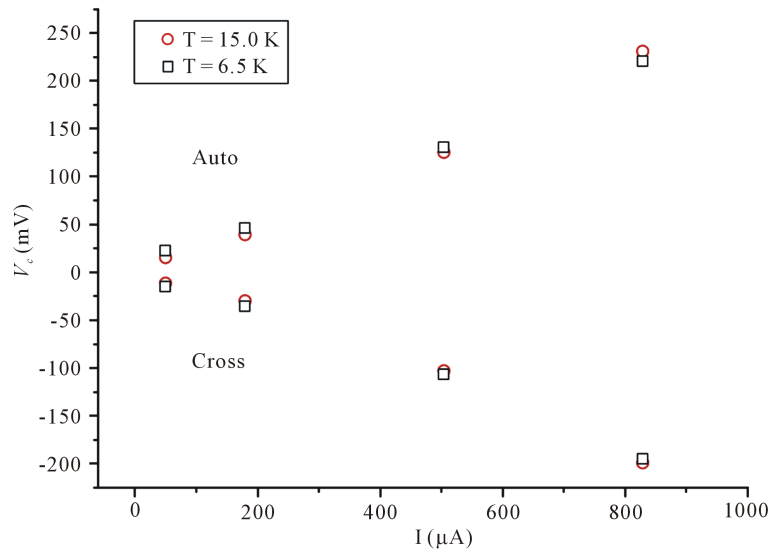

**Figure 2.** Transport fluctuation correlation at 6.5 K (squares) and 15.0 K (circles) as a function of the incident beam intensity ($I = I_\text{eff}$) in a Y-shaped Nb film of small resistance (R = 200 Ohm, at 293 K) and superconductive transition at $T_c = 9.2$ K. Current independent fluctuations, such as thermal noise and residual electronic noise, have been subtracted from the readings of $V_C$. The vertical size of the squares represents the experimental error on $V_C$. The slope of both curves yields $(t_a \, t_b) \simeq 0.25$.

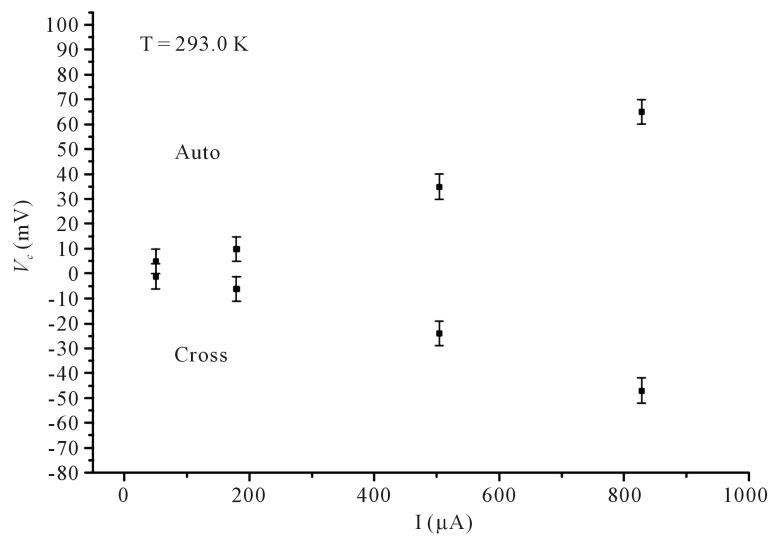

**Figure 3.** As in **Figure 2**, at $T$ = 293 K. The slope of the curves yields $\left(t_a\ t_b\right) \simeq 0.06$.

All the specimens showed up the same behaviour; and **Figure 2** shows the results recorded with the 20Ω-specimen at $T$ = 15.0 K ($>T_c$ = 9.2 K) and $T$ = 6.5 K ($<T_c$). At $T$ = 15.0 K, the negative slope and value of the cross-correlation as well as the positive slope and value of the auto-correlation, expected for Fermi particles according to formulae (1), are clearly recorded. At $T$ = 6.5 K, the sign and the value of the measured slop are the same as those above $T_c$, as predicted by formulae (2) for condensation of the Cooper pairs. At these low temperatures, below or little above $T_c$, the measured slope of the correlations yields a value of $t_a\ t_b \simeq 0.25$, in good agreement with the reasonable prediction of $t_a = t_b \simeq 50\%$. At much higher temperatures, the thermal noise reduces both the cross-correlation and the auto-correlation, as it is shown in **Figure 3**[1].

## 4. Conclusion

We have performed an experiment whose results give direct evidence of the statistical nature of the electron gas in Nb films, at temperatures above $T_c$ (fermions) and at temperatures below $T_c$ (Cooper pairs in a single state of energy and momentum). The conceptual scheme of the experiment, which measures the intensity-fluctuation correlation of two beams of particles partitioned off at a *Y*-shaped obstacle, is the HBT (Hanbury Brown and Twiss) scheme, here adapted to the electron gas in a conductor. For lack of availability of proper samples, we could not perform an analogous experiment with high-temperature superconductors.

---

[1]We may note that the temptation to find an approach to our experiment, alternative to the one presented here, in terms of the Ising model would not appear fruitful. We have measured correlations between one-dimensional electric currents near the quantum superconducting transition of the electric gas in Nb films. Vice versa, the Ising model is a lattice-based classical model whose one-dimensional version has no phase transition. The application of the two-dimensional version (Onsager solution) would appear inappropriate to represent our experimental scheme, and in any case it would require the difficult task to include quantum effects, "lattice-gas" sites, momentum of charged particles.